\documentclass{cernyrep}
\usepackage{amsmath}
\usepackage{siunitx}

\usepackage[T1]{fontenc}
\usepackage[bookmarks, colorlinks=true, linktoc=page, pdftex, linkcolor=black, citecolor=black, urlcolor=blue]{hyperref}

\usepackage{fancyhdr}
\fancyhfoffset{4 mm}
\fancypagestyle{ARTTITLE}{%
\fancyhf{}
\lhead{\raggedright {\it Mechanical \& Materials Engineering for Particle Accelerators and Detectors}\raggedright \\ CERN Accelerator School Proceedings ---  Sint-Michielsgestel,  Netherlands, 2024\hfill}
\lfoot{\hspace{3mm} Available online at \url{https://cas.web.cern.ch/previous-schools}}
\rfoot{\thepage\hspace*{3mm}}

}

\usepackage{graphicx}
\usepackage{booktabs}
\usepackage{siunitx}
\usepackage{float}
\usepackage{caption}
\usepackage{mdframed}
\usepackage{enumitem}
\usepackage{afterpage}
\usepackage{lipsum}
\usepackage{array}
\usepackage{xcolor, soul}
\usepackage{colortbl}
\usepackage{wrapfig}
\usepackage[OT2,OT1]{fontenc}

\newmdenv[backgroundcolor=lightgray]{importantbox}
\providecommand*{\Eq}[1][s]{\ifthenelse{\equal{#1}{b}}{Equation}{Eq.}}
\providecommand*{\Eqs}[1][s]{\ifthenelse{\equal{#1}{b}}{Equations}{Eqs.}}
\providecommand*{\Figure}[1][s]{\ifthenelse
                                      {\equal{#1}{b}}{Figure}{Fig.}}
\providecommand*{\Figures}[1][s]{\ifthenelse
                                      {\equal{#1}{b}}{Figures}{Figs.}}
\providecommand*{\Ref}[1][s]{\ifthenelse{\equal{#1}{b}}{Reference}{Ref.}}
\providecommand*{\Refs}[1][s]{\ifthenelse{\equal{#1}{b}}{References}{Refs.}}
\providecommand\Table{Table}
\providecommand\Tables{Tables}

\providecommand\Section{Section}
\providecommand\Sections{Sections}

\providecommand*\Eref[2][s]{\Eq[#1]~(\ref{#2})}

\providecommand*\Fref[2][s]{\Figure[#1]~\ref{#2}}

\providecommand*\Sref[1]{\Section~\ref{#1}}

\begin{document}

\title{Colliders}
\author{H.~Schmickler}
\institute{CERN, Geneva, Switzerland}

\begin{abstract}
Modern particle physics relies on high energy particle accelerators to provide collisions of various types of elementary particles in order to deduce fundamental laws of physics or properties of individual particles. The only way to generate particle collisions at extremely high energies is to collide particles of counter-rotating beams...called "particle-colliders". 

This write-up gives a short briefing on the physics motivation of various particle colliders ($e^+e^-$ colliders, $pp$ colliders,  ...), a summary of the historical evolution and a mathematical treatment to describe collider performance.

\end{abstract}


\maketitle
\thispagestyle{ARTTITLE}

\section{Introduction}
The "heaviest" constituents of the Standard-Model have masses of about 100 GeV/$c^2$. If such objects should be produced in particle collisions, the center of mass energy (cms energy) of the two colliding particles must be of that order or larger.
In particle physics we distinguish two types of experimental set-ups:
\begin{itemize}
\item{fixed target experiments: A particle beam hits after acceleration a target at rest.
}
\item{collision experiments: Two counter rotating particle beams are both accelerated and brought into collision. The experimental detector encloses the collision point.
}
\end{itemize}

The kinematics of a particle with mass $m$ can be expressed 
by its momentum $\vec{p}$ and energy $E$ which form a
four-vector {\bf{p}}$~=~(E,\vec{p})$. The
square {\bf{p}}$^{2}$ is (with the redefinition of units by setting the~speed of light to 1) (see for example: \cite{bib:COL1})
\begin{equation}\label{eq:002}
     {\bf{p}}^{2}~~=~~E^{2} - \vec{p}^{2}~=~m^{2}
\end{equation}
~\\
In the collision of two particles of masses $m_{1}$ and $m_{2}$ the
total centre of mass energy can be expressed in the form
\begin{equation}\label{eq:003}
     ({\bf{p}}_{1} + {\bf{p}}_{2})^{2}~=~E^{2}_{cm}~=~(E_{1} + E_{2})^{2} - (\vec{p}_{1} + \vec{p}_{2})^{2}
\end{equation}
~\\
This is the available energy for physics experiments.
~\\
In the case of a collider where the collision point is at
rest in the laboratory frame (i.e. $\vec{p_{1}}$ = $-\vec{p_{2}}$),
the~centre of mass energy becomes:
\begin{equation}\label{eq:004}
     E^{2}_{cm}~=~(E_{1} + E_{2})^{2}
\end{equation}
~\\
When one particle is at rest, i.e. in the case of so-called fixed
target experiments, (i.e. $\vec{p_{2}}$ = 0), we get:
\begin{equation}\label{eq:005}
     E^{2}_{cm}~=~(m_{1}^{2} + m_{2}^{2} + 2m_{2}E_{1,lab})
\end{equation}
A comparison for different types of collisions is made in Table \ref{tab:01}.
\begin{table}[htb]
\centering
\caption{Centre of mass energy for different types of collisions.}\label{tab:01}
\begin{tabular}{|c|c|c|}  \hline \hline
    &        &     \\
          & \textbf{$\mathbf{E}_{\mathrm{cm}}$ as collider (GeV)} & \textbf{$\mathbf{E}_{\mathrm{cm}}$ with fixed target (GeV)} \\
    &        &     \\
\hline
    &        &         \\
p on p  (7000 on 7000 GeV)   &  14000          & 114.6           \\
    &        &     \\
e on e   (100 on 100 GeV)   &  200          & 0.32            \\
    &        &     \\
e on p   (30 on 920 GeV)   &  235          & 7.5            \\
    &        &     \\
\hline \hline
\end{tabular}
\end{table}
From this table it is rather obvious why colliding beams are
necessary when centre of mass energies of the order of 100 GeV are required for
particle physics experiments. In fixed target experiments most of the beam energy is "wasted" to accelerate the center of mass and not available for new particle production.
Nevertheless most particles of our present Standard model (all leptons, all quarks other than the t-quark, neutrinos) have been discovered in fixed target experiments.

\section{Main characteristics of a collider}
\label{sec:COL-CHAR}
\begin{figure}[ht]
  \centering
  \includegraphics[width=0.5\textwidth]{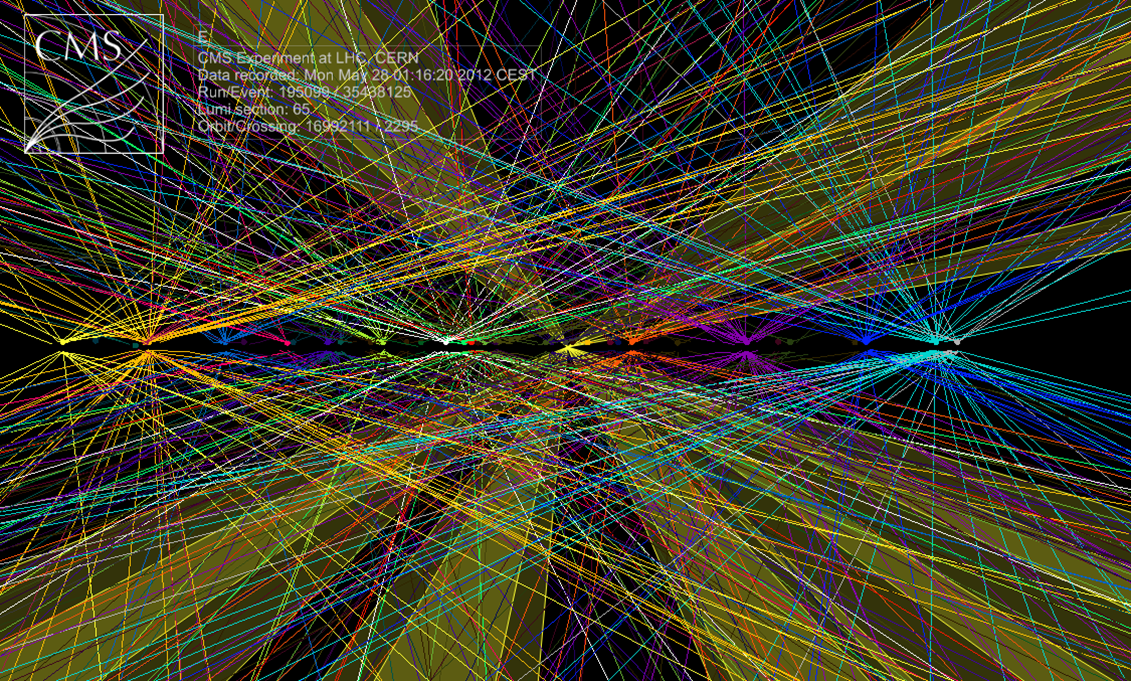}
  \caption{Measured charged tracks close to the vertex region of a collider. The tracks coming from individual particle interactions are coloured in different colors.}
  \label{fig:COL2}
\end{figure}

Particle-Colliders are very expensive and complex installations and up to date only a few "hand-fulls" have been constructed and operated.
The following criteria are useful to classify them:
\begin{itemize}

\item{ 
\textbf{Accelerated particles}: Particles must be charged for acceleration, apart from that leptons and their antiparticles, proton and their antiparticles or ions can be accelerated in a collider. So all but one colliders that have been built are synchrotrons, in which the colliding beams are accelerated and then kept at constant energy for several hours.
After every revolution the beams encounter themselves and produce particle collisions.
In case of a particle/antiparticle collider it is sufficient to construct a synchrotron with a single beam tube, since the Lorentz-force keeping the particles on a bent orbit changes sign with the change of particle charge. Hence both counter-rotating beams can be kept in the same ring. Special care has to be taken such that in the case of several bunches the bunches do only collide in the wanted interaction regions.
If particles collisions of the same type (for example proton-proton) are wanted, the accelerator must have two beam pipes, which are made to intersect in the interaction regions. There the beams will interact with a small crossing angle.

So far proton-proton, proton-antiproton, electron-positron and various ion species colliders have been built. Also lepton-hadron colliders have been built in order to probe with the elementary electrons the internal structure of the protons.
}

\item{
\textbf{linear - or circular colliders:}
In the case of a linear collider one uses two independent linac structures in order to accelerate the two beams and makes them collide in a central region. The~beams are only used once for collisions and are disposed directly after the collision point.
Such colliders are only envisaged for the creation of highest particle masses (above 100 GeV/$c^2$). The~linear structures will avoid energy loss due to synchrotron radiation on a curved orbit. So far only one linear collider has been built (SLC:= Stanford Linear Collider), which used in a clever way one linac to accelerate both electrons and positrons plus at the end of the acceleration stage two $180^0$ return arcs in order to make the two beams collide head-on.
}
\item{
\textbf{maximum collision energy:}
In the case of a circular collider the acceleration of the beams can be stopped at any moment, so in principle collision energies from twice the injection energies to twice the maximum synchrotron energies are possible. In a linear collider the beams have to pass in any case through the whole linac structure, which in general can only be operated for a fixed energy. Changes of the energy require significant changes to the settings of the accelerators if not changes to the hardware layout.
}
\item{
\textbf{interaction rate:}
A very important quality indicator of a collider is the rate of particle interactions per unit time. This is characterized by a proportionality factor between the interaction rate $\dot N$ and its cross-section $\sigma_{\textit{int}}$
\begin{equation}
\dot N \ =\ \cal{L}\cdot \sigma_{\textit{int}} 
\end{equation}
which is called the luminosity $\cal{L}$ of the collider. The luminosity depends on the intensity of the~accelerated beams, on the frequency by which they collide and on the density of the beams at the~collision point.
Standard units for the cross section are "barns". 1 barn [b] = $10^{-24}\ cm^2$. With this description one imagines the probability for an interaction as a small surface (measured in barns), which needs to be hit by the projectile. Accordingly the luminosity is given in units of $cm^2/s^{-1}$, so the product $\cal{L}\cdot \sigma$ has $s^{-1}$ (a counting rate) as unit.
}
\item{
\textbf{detector occupancy or pile-up:} Another performance criterion has come up in recent years in the~LHC (pp-collisions at 13 TeV/$c~2$). In the case of the LHC the total cross-section of the pp interaction and the luminosity of the collider are so large, that the production rate gives more than one beam interaction per beam crossing. Actually beam crossings with more than 100 proton-proton interactions have been observed in the LHC. At these energies the average charged multiplicity of the collison-secondaries is also about 100, so beam crossings with up to 10.000 charged tracks in a~detector have been measured. Hence the detector performance puts an upper limit to the luminosity that should be produced; otherwise the collisions can not be measured correctly. The~number of interactions per beam crossings is called "detector-occupancy". If the collider produces a too high occupancy, the luminosity has to be lowered artificially. A nice example of a collision snapshot with about 10 interactions per beam crossing is shown in \Fref{fig:COL2}.
}
\end{itemize}

\begin{figure}[ht]
  \centering
  \includegraphics[width=0.9\textwidth]{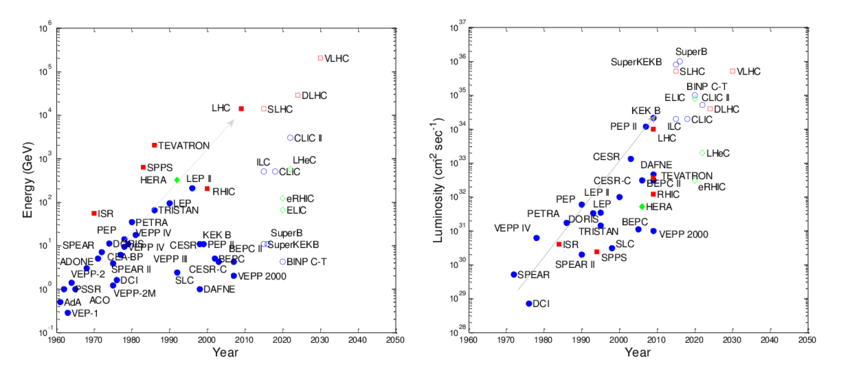}
  \caption{Center of mass energy (left) and luminosity (right) of many particle accelerators as a function of time (taken from \cite{bib:COL2}.
Empty symbols are for constructed projects, empty symbols are planned projects
(blue for $e^+ e^-$, red for pp and green for ep colliders).}
  \label{fig:COL1}
\end{figure}

Figure~\ref{fig:COL1} summarizes the present situation with an overview of constructed and planned collider projects worldwide from the year 1960 to today. It is interesting to see the increase in centre-of-mass energy over the years and also the increase in luminosity (see \Sref{sec:lumi}).

\newpage
\section{What type of particles for the collisions?}

The choice of particles for the collisions will of course be dominated by the research objectives of the installation. The following sections explain some of the physics objectives and the conceptually necessary technology issues for all built or envisaged particle colliders:

\subsection{Electron-positron collisions ($e^+e^-$):}

The lightest and stable members of the lepton family, electrons and positrons are considered elementary, point-like particles, which interact through the electromagnetic and the weak force. This makes them the ideal candidates as choice for collisions, most importantly because of the elementary nature of the~particles one knows the interaction center of mass energy by knowing the two beam energies. Precisions of $10^{-5}$ for the collision energy have been obtained, which has allowed to determine the masses of the~created particles to that precision \cite{bib:COL3}. The cross sections for $e^+e^-$ collisions is low in the range of nano-barns to femto-barns, which yields even with the highest available luminosities event rates for which one can easily construct detectors.
Electrons are very easy to produce and accelerate, positrons need to be produced by interactions of an electron beam on a target and later reduction of transverse momenta in damping rings.
The main issue preventing  $e^+e^-$ colliders to be the only research tool is synchrotron radiation. Every charged particle following a bent orbit radius $\rho$ looses energy by emission of a photon (:= synchrotron radiation), which can be characterized by the following two quantities:\\
critical energy $E_c$:
\begin{equation}
E_c\ =\ \frac{3}{2}\hbar c \frac{\gamma ^3}{\rho}.
\label{eq:toto1}
\end{equation}
Energy loss per turn $U_0$
\begin{equation}
 U_0\ = \ \frac{4\pi \alpha\hbar c \gamma ^4}{3\rho}
 \label{eq:toto2}
\end{equation}
with values of the momentum compaction $\alpha \simeq 10^{-4}$. The photon energy increases with $\gamma ^3$ and the energy loss in synchrotron radiation over one turn as $\gamma ^4$. At beam energies around 100 GeV corresponding to a Lorentz factor of
$\gamma \simeq 2\cdot 10^5$ 3\% of the particle energy was lost in LEP in a single turn ($\rho \simeq 7km$). With protons having a rest mass 1836 times higher than electrons, the Lorentz factor is correspondingly smaller and synchrotron radiation only becomes relevant in proton-proton collisions at TeV energy scales.
So in practical terms for $e^+e^-$ collision energies well above 200 GeV the only alternative seems to be linear colliders.
Details on the last built and operated collider LEP and on the two linear collider projects ILC or CLIC can be found in the last section.

\subsection{Muon collisions ($\mu ^+\mu ^-)$:}

There have also been studies for muon-colliders \cite{bib:COL4},\cite{bib:muon}. Muons could be generated in the decay channel of pions, then "quickly" accelerated within the muon life time and made collide in a reasonably sized collider ring. 
Like electrons, muons are also elementary particles, but their restmass is about 200 times bigger than the restmass of electrons. Referring to \Eref{eq:toto2} one can see that the energy loss per turn becomes even at TeV-range energies insignificant ($\gamma^2$ scales like $1/m$), hence synchrotron radiation losses will not be a limiting factor for a muon collider.
As another advantage one can count the fact that the cross-section for producing the vector-boson $Z^0$ in lepton collisions scales with $m^2$. So even with much lower luminosity a "$Z^0$-factory" could be imagined with a muon collider.

For the design of a muon collider the fundamental problems come from the short life-time of muons.
At low energy the average life-time of a muon beam would just be 2.2 [$\mu$s], so before being able to produce physics events the muon beam would already have decayed into electrons/positrons and their accompanying neutrinos.

\begin{figure}[ht]
  \centering
  \includegraphics[width=0.5\textwidth]{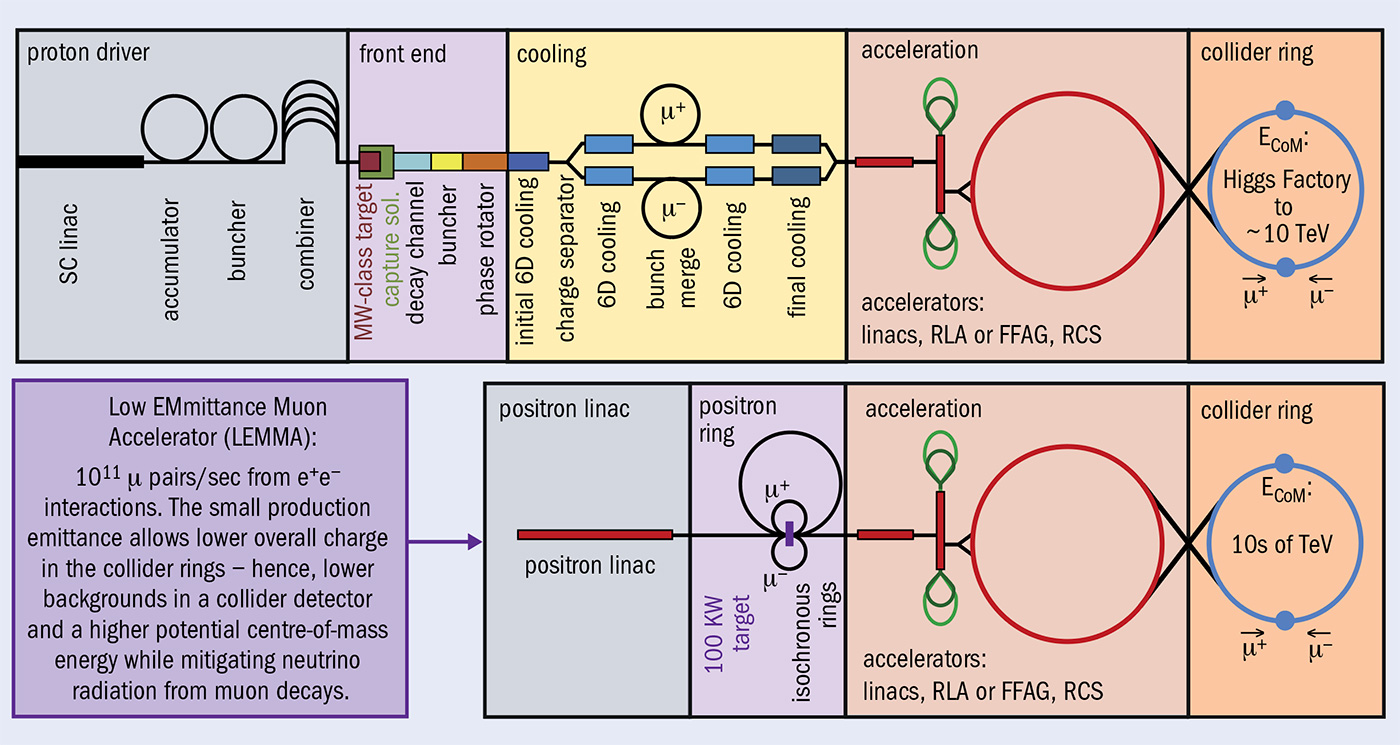}
  \caption{Top: Schematic layout of a potential muon collider with a muon source based on protons. Bottom: Schematic layout of a potential muon collider with a muon source based on positrons (taken from \cite{bib:muon}).}
  \label{fig:toto3}
\end{figure}

Based on that fact one can list the conceptually biggest problems for the design of a muon collider:

\begin{itemize}
\item{
The muon beams, created after proton-matter collisions or even positron-matter interactions and the following decay of pions into muons (see \cite{bib:muon}) need to be accelerated extremly fast in order prolong their life-time due to the rapid increase of the Lorentz-factor. A combination of a linac and RFQs is envisaged.
}
\item{
The created muon beam has a very large transverse emittance, so cooling is necessary in order to get particle densities high enough to make collisions worth while.
The classical damping ring approach is impossible due to the short lifetime of the muon beam. The presently envisaged solution is to reduce the transverse emittance during acceleration by making the beams pass once through absorber material (reducing the overall momentum and adding new momentum only in the direction of flight by the accelerating cavities. A process similar to radiation damping of electrons/positrons at high energies in the presence of synchrotron radiation.
}
\item{
During the collisions phase the entire particle beams decays into electrons/positrons and neutrinos and the electrons/positrons will depose all their energy into the accelerator structures (vacuum-tube and magnets). So in particular in the case of superconducting magnets significant shielding/absorbers needs to be provided in order to avoid save accelerator operation.
}
\end{itemize}
Figure~\ref{fig:toto3} shows the principal layout of a the presently envisaged muon-collider
concepts.

\subsection{Proton-proton collisons ($pp$ or $p\bar p$):}

Almost free of synchrotron radiation protons can be accelerated in a circular accelerator to energies in the~multi-TeV range. Historically also proton-antiproton colliders have been built, but due to the~difficulties in getting large stacks of dense antiproton beams, modern designs for highest collison energies do not consider proton-antiproton colliders anymore due to the lower achievable luminosities \cite{bib:COL5}.

\begin{wrapfigure}[17]{r}{0.5\textwidth}
  \begin{center}
	\includegraphics[width=0.32\textwidth]{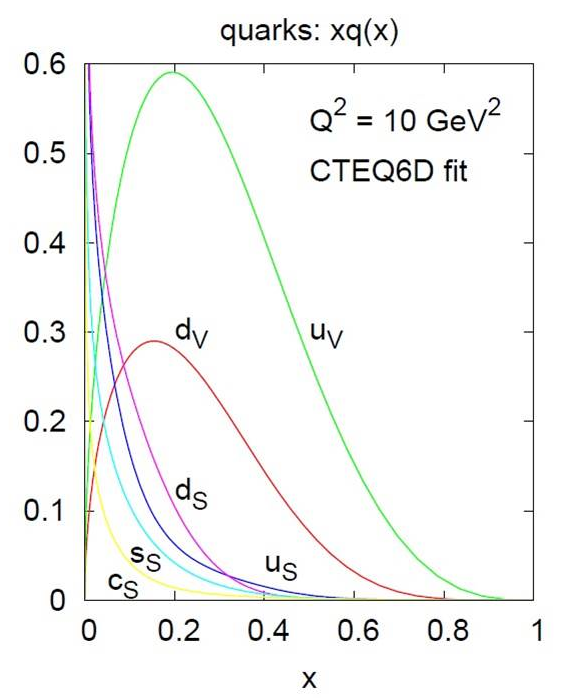}
  \end{center}
	\caption{modelled momentum distribution of valence- (index v) and sea-quarks (index s)}
	\label{fig:COL5}
\end{wrapfigure}

\noindent Protons are not elementary particles. In a simplistic way one can describe them as three valence quarks, a "sea" of quark-antiquark pairs and gluons, which are the conveyors of the strong force (more detailed reading for example in  \cite{bib:COL6}).  We retain here that the valence quarks carry most of the~energy of the proton and as the illustration on the~left side shows they carry on average 20\% of the~proton energy with a wide probability distribution.
An important consequence is that in proton-proton collisions one does not know the~initial energy of the collision partners. Only in some cases the initial energy can be reconstructed by summing up the measured energies of all secondary particles. In general the lack of knowledge of the~initial energy is a significant limitation in the~data analysis of proton-proton collisions.

\vspace{1cm}
 
\begin{wrapfigure}[22]{r}{0.5\textwidth}
  \centering
  \includegraphics[width=0.5\textwidth]{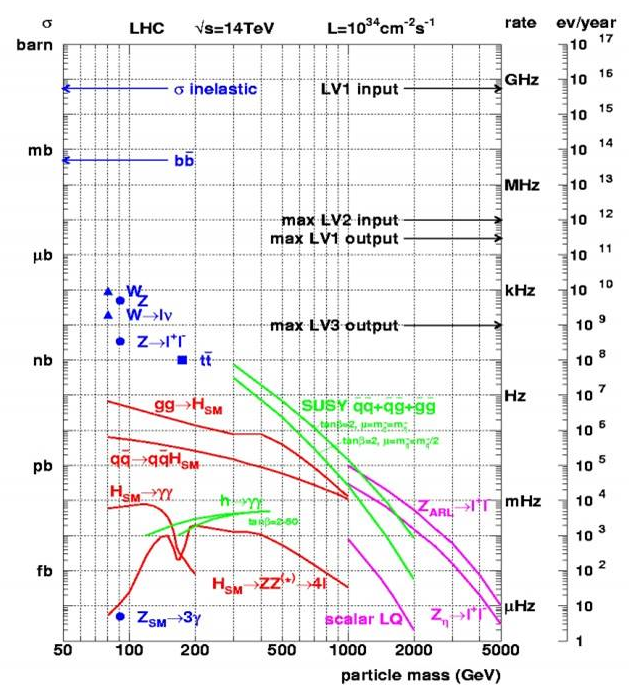}
  \caption{cross-sections for various interesting physics processes as function of the created particle mass. The scales on the right indicate the event rate for a luminosity of $10^{34}cm^{-2}s^{-1}$.}
    \label{fig:COL6}
\end{wrapfigure}

\noindent A very significant difference between $pp$ collisons and $e^+e^-$  collisons is the high total cross-section of $pp$ collisons. At 13 TeV cms energy the~cross-section is about 100 [mb], which is due to the~strong interaction $10^5$ times higher than in the~case of the electro-weak interactions in $e^+e^-$ collisons.
But most of these many $pp$ interactions only create secondary fragments, which are of little interest to study. Only a very small fraction of the produced interactions represents an event of rare particle production. The situation is depicted in 
\Fref{fig:COL6}.

One clearly sees with the high total cross-section ($\sigma_{\rm{inelastic}}$) event rates close to 1 GHz are possible, which explains the large detector occupancy at 40 MHz bunch crossing frequency of the~LHC (see also \Sref{sec:COL-CHAR}).
Such large amounts of data cannot be recorded. With a sophisticated system of trigger electronics, a subset of events with interesting content is preselected for recording.
As you can see the rate for production a~Higgs-boson through gluon fusion ($gg\rightarrow H_{SM}$) is only about 1 Hz.

\subsection{Ion colliders}
In terms of technology ion colliders do not differ much from proton colliders, in principle all ion colliders can also accelerate and collide protons.
The main difference is in the research objectives and the correspondingly different design of the physics detectors. Ions being composed of many protons and neutrons have a more complex internal structure, so in the event of an ion-ion interaction very high densities of secondary particles are created. 

One of the largest presently operating ion collider is RHIC (= Relativistic Heavy Ion Collider) on Long Island (US).
Several different species of ions can be accelerated and collided in two superconducting rings. More details on RHIC can be found under \cite{bib:COL7}.
The LHC at CERN is also been used to collide ion-beams during short periods at the end of each running period.

\subsection{Electron-proton- (ep) colliders}
At DESY close to Hamburg (GE) a large collider for protons and electrons has operated in the years 1992 - 2007 for research operation. It used a normal conducting ring for the acceleration of electrons (and occasionally positrons) up to 40 GeV and a superconducting ring for the acceleration of protons to 820 GeV. In these electron-proton collisions, the point-like electron acts like a tiny probe that scans the inside of the proton and reveals its inner structure. The higher the energy of the particle collision, the deeper physicists are able to gaze into the proton, and the more insights they obtain about the inner structure of the proton and the fundamental forces of nature. So called structure functions as shown in \Fref{fig:COL5} are based on measurement results of ep interactions.
More details can be found at \cite{bib:COL8}.

\subsection{Photon-photon collisions}
The collisions of photons sounds crazy at the first sight, since photons as neutral particles can not be accelerated in an accelerator.
But synchrotron radiation "helps": Close to the interaction-point of electrons and positrons the Coulomb-force bends the particles towards each other, which stimulates the emission of photons. In some cases a photon is emitted from an electron and a second photon from a positron and they can interact. So photon-photon collisions are a subset of $e^+e^-$ collisions and they provide a~large field of research. Although the collisions of the electrons and positrons are always at the sum of the~beam energies, the photon energies represent a large spectrum and hence the collision products can have a large energy spectrum (more reading in \cite{bib:COL9}).

\subsection{Wakefield accelerators}
All accelerators described so far use conventional Rf-Cavities for the particle acceleration. Accelerating gradients of up to 100 MV/m can be reached in extreme cases.
Since many years people try to use wakefields in plasma for the acceleration of electrons or positrons. Theoretically much higher gradients of up to 100 GV/m could be reached.
A full lecture in the present CAS-course is dedicated to this research \cite{bib:COL10},\cite{bib:COL11}.

\subsection{Summary}

In summary one should highlight the major aspects of the most important collider types built so far:
\begin{itemize}
\item{
\textbf{$e^+e^-$ colliders}\\
The highest collison energies produced were just above 200 GeV in LEP. Presently people study a project of a circular collider (FCC-ee) up to 360 GeV and two linear colliders up to a maximum of 3 TeV of beam energy. Such colliders produce almost mono-energetic collisions of elementary particles with very clean experimental background conditions
(no hadronic interactions).
$e^+e^-$ colliders are often quoted as "instruments for precision measurements".
}
\item{
\textbf{$pp$ colliders}\\
Presently the LHC produces $pp$-collisions at 13 (max. 14) TeV. There is a project to construct a pp-collider at 100 TeV (FCC-pp). $pp$-colliders produce the interaction of composite particles, in which the interacting constituents have a large energy spectrum.
Most of the interactions are "hadronic debris", which obscures somehow the analysis of the underlying interesting physics events. The data analysis in $pp$ collisions is much more involved, but the research potential is very large, since presently only this way the highest collision energies can be produced.
Many, if not most, of new particle discoveries are based on the observation of $pp$ or $p\bar{p}$ collisons.
}
\item{
\textbf{ion colliders}\\
In terms of design and construction ion colliders are similar to $pp$ colliders, although there are major differences in achievable bunch intensities and luminosities.
In most cases ion collisions are part of the experimental program of a collider, which also is used for $pp$ collisions.
The physics research goals of ion colliders are different, a fundamental interest is to use the high particle density at collision to emulate matter conditions as expected during the big bang at the creation of our universe.
}
\item{
\textbf{$ep$ colliders}\\
Only few of these colliders have been built in the past (HERA the most important example), but there are also proposals to complement the LHC with an electron linac in order to study $ep$-collisons at very high energies.
The research potential for new discoveries is limited on such colliders, but insights into the constituent particles of hadronic matter is best obtained with such colliders.
}
\end{itemize}

\section{Computation of luminosity}
\label{sec:lumi}
\subsection{Fixed target luminosity}
\begin{figure}[ht]
\begin{center}
\includegraphics[width=0.5\textwidth]{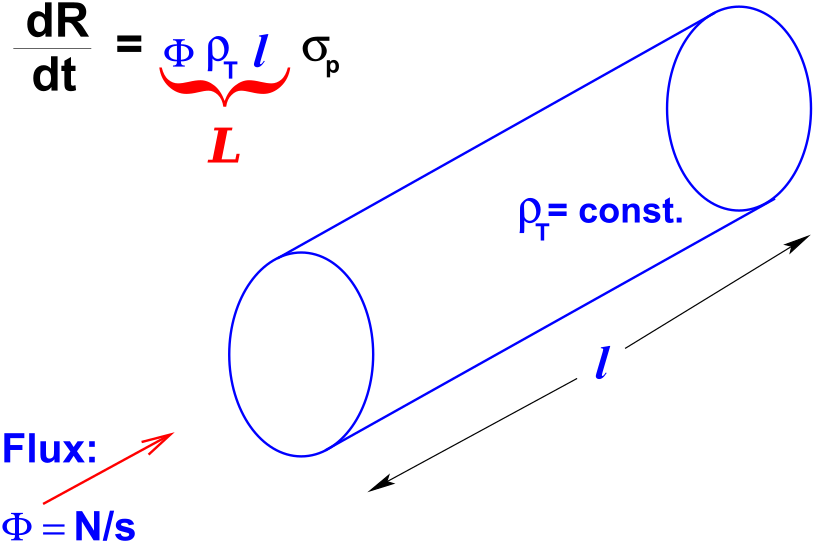}
\end{center}
\caption{\it{Schematic view of a fixed target collision.}}\label{fig:01}
\end{figure}

In order to compute a luminosity for fixed target
experiment, we have to take into account the properties of both,
the incoming beam and the stationary target.
The basic configuration is shown in Fig.~\ref{fig:01}.
The~incoming beam is characterized by the flux $\Phi$, i.e. the number
of particles per second.
When the~target is homogeneous and larger than the incoming beam,
the distribution of the latter is not important for the~luminosity.

The target is described by its density $\rho_{T}$ and its length $l$.
With a definition of the luminosity like:
\begin{equation}\label{eq:006}
     {\cal{L}}_{FT}~=~\Phi \rho_{T} l                           
\end{equation}
we write the interaction rate
\begin{equation}\label{eq:007}
     \frac{dR}{dt}~~=\Phi \rho_{T} l~\cdot\sigma_{p}~=~{\cal{L}}_{FT}~\cdot~\sigma_{p}
\end{equation}
as desired.
\subsection{Colliding beams luminosity}
In the case of two colliding beams, both beams serve as target and 
"incoming" beam at the same time.
Obviously the beam density distribution is now very important and the
generalization of the above
expression leads to the convolution of the 3-D distribution functions.
\begin{figure}[ht]
\begin{center}
\includegraphics[width=0.5\textwidth]{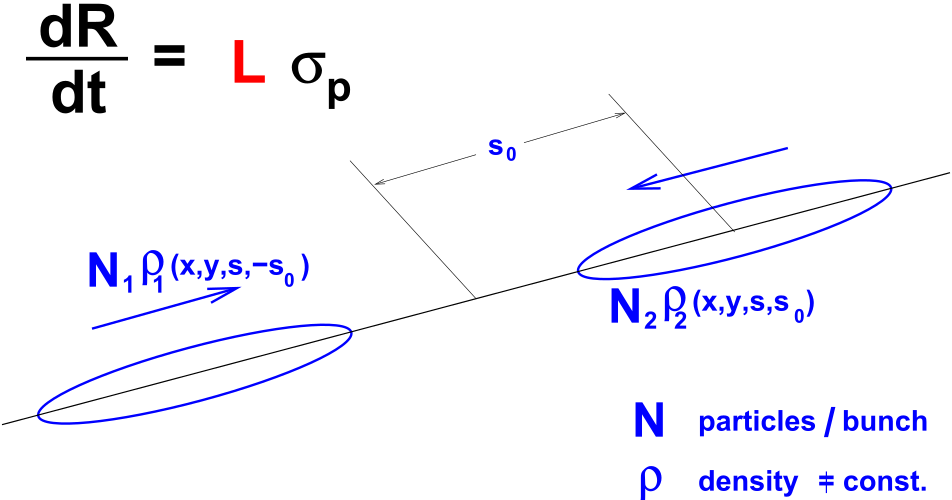}
\end{center}
\caption{\it{Schematic view of a colliding beam interaction.}}\label{fig:02}
\end{figure}
We treat the case of bunched beams, but it can easily be extended
to unbunched beams or other concepts like very long bunches.
A schematic picture is shown in Fig.\ref{fig:02}.
Since the two beams are not stationary but moving through each other,
the~overlap integral depends on the longitudinal position of
the bunches and therefore on the time as they move towards and through
each other.
For our integration we use the distance of the two beams to the central
collision point $s_{0} = c\cdot t$ as the "time" 
variable (see Fig.\ref{fig:02}).
A priori the two beams have different distribution functions and different
number of particles in the beams.
~\\
The overlap integral which is proportional to the luminosity {\cal{L}}
we can then write as:
\begin{equation}\label{eq:008}
{{\cal L} \propto {{K}} \cdot \int\int\int\int_{-\infty}^{+\infty}} \rho_{1}(x,y,s,-s_{0}) \rho_{2}(x,y,s,s_{0}) dxdydsds_{0}. 
\end{equation}
Here $\rho_{1}(x,y,s,s_{0})$ and $\rho_{2}(x,y,s,s_{0})$ are the
time dependent beam density distribution functions.
We assume, that the two bunches meet at $s_{0}~=~0$.
Because the beams are moving against each other, we have to
multiply this expression with a kinematic factor \cite{kinfact}:
\begin{equation}\label{eq:009}
 {{K}} = \sqrt{(\vec{v_{1}} - \vec{v_{2}})^{2}  -  (\vec{v_{1}}\times\vec{v_{2}})^{2}/c^{2}}.
\end{equation}
~\\
In the next step we assume head-on collisions ($\vec{v_{1}} = -\vec{v_{2}}$)
and that all densities are uncorrelated in all planes.
In that case we can factorize the density distributions and get for
the overlap integral:
\begin{equation}\label{eq:010} 
~~~~~~{\cal L} = {{2}} N_{1}N_{2}fN_{b}
\displaystyle{\int\int\int\int_{-\infty}^{+\infty}} 
\rho_{1x}(x)\rho_{1y}(y)\rho_{1s}(s - s_{0})
\rho_{2x}(x)\rho_{2y}(y)\rho_{2s}(s + s_{0})~~dxdydsds_{0}.
\end{equation} 
~\\
We have completed the formula with the beam properties necessary to 
calculate the value of the luminosity:
$N_{1}$ and $N_{2}$ are the intensities of two colliding
bunches, $f$ is the revolution frequency and $N_{b}$ is the number
of bunches in one beam.

To evaluate this integral one should know all distributions.
An analytical calculation is not always possible and a numerical
integration may be required.
However in many cases the beams follow "reasonable" profiles
and we can obtain closed solutions.
\subsection{Luminosity of Gaussian beams colliding head-on}
Often it is fully justified to assume Gaussian distributions.
The luminosity is determined by the overlap of the core of
the distributions and the tails give practically no
contribution to the luminosity. We shall come back to this
point in a later section.

For the first calculation we assume Gaussian profiles in
all dimensions of the form:
\begin{equation}\label{eq:011}
\ \ \displaystyle{\rho_{iz}(z) = \frac{1}{\sigma_{z}\sqrt{2\pi}}
\exp{\left(-\frac{z^{2}}{2\sigma_{z}^{2}}\right)}}
\ {\mathrm{where}} \ \ i = 1,2,\ \  z = x,y
\end{equation}
\begin{equation}\label{eq:012}
\ \ \displaystyle{\rho_{s}(s \pm s_{0}) =
\frac{1}{\sigma_{s}\sqrt{2\pi}}
\exp{\left(-\frac{(s \pm s_{0})^{2}}{2\sigma_{s}^{2}}\right)}}.
\end{equation}
~\\
Furthermore we assume equal beams, i.e.:
{{$\sigma_{1x} = \sigma_{2x}, \sigma_{1y} = \sigma_{2y},\sigma_{1s} = \sigma_{2s}$}}. \\
~~~\\
Next we assume the number of particles per bunch $N_{1}$ and $N_{2}$, 
a revolution frequency of $f$ and the number of bunches we call $N_{b}$.
In the case of exactly head-on collisions of bunches travelling almost
at the speed of light, the kinematic factor becomes 2.

Using this in Eq.~(\ref{eq:010}) we get the first integral:
\begin{equation}\label{eq:013} 
{\cal L} = \frac{2\cdot N_{1}N_{2}fN_{b}}{(\sqrt{2\pi})^{6}\sigma_{s}^{2}\sigma_{x}^{2}\sigma_{y}^{2}}
\displaystyle{\int\int\int\int
e^{-\frac{x^{2}}{\sigma_{x}^{2}}}
e^{-\frac{y^{2}}{\sigma_{y}^{2}}}
e^{-\frac{s^{2}}{\sigma_{s}^{2}}}
e^{-\frac{s_{0}^{2}}{\sigma_{s}^{2}}}  dxdydsds_{0} }\\
\\
\end{equation}
integrating over $s$ and $s_{0}$, using the well known formula:
\begin{equation}\label{eq:014} 
 \displaystyle{\int^{+\infty}_{-\infty} e^{-a t^{2}} dt~~~=~~~\sqrt{\pi/a}}
\end{equation}
we get a first intermediate result:
\begin{equation}\label{eq:015} 
{\cal L} = \frac{2\cdot N_{1}N_{2}fN_{b}}{8(\sqrt{\pi})^{4}\sigma_{x}^{2}\sigma_{y}^{2}}
\displaystyle{\int\int 
e^{-\frac{x^{2}}{\sigma_{x}^{2}}}
e^{-\frac{y^{2}}{\sigma_{y}^{2}}} dxdy }\\
\\
\end{equation}
finally, after integration over x and y:
\begin{equation}\label{eq:016} 
{{\Longrightarrow}}~~~{{{\cal L} = \displaystyle{\frac{N_{1}N_{2}fN_{b}}{4\pi\sigma_{x}\sigma_{y}}}}}.
\end{equation}
This is the well-known expression for the luminosity of two Gaussian beams
colliding head-on.
It shows how the luminosity depends on the number of particles per bunch and
the beam sizes.
This reflects the~2-dimensional target charge density we have seen in 
the evaluation of the fixed target luminosity.
~\\
For the more general case of: {{$\sigma_{1x} \neq \sigma_{2x}, \sigma_{1y} \neq \sigma_{2y}$, but still assuming approximately equal bunch lengths $\sigma_{1s} \approx \sigma_{2s}$}} we get a modified formula:
\begin{equation}\label{eq:017} 
{{ {{{\cal L} = \displaystyle{\frac{N_{1}N_{2}fN_{b}}{2\pi\sqrt{\sigma^{2}_{1x} + \sigma^{2}_{2x}}\sqrt{\sigma^{2}_{2y} + \sigma^{2}_{2y}}}}}}  }}.
\end{equation}
~\\
This formula is easy to verify and also straightforward to extend to 
other cases.
Here it is worth to mention that the luminosity does not depend on
the bunch length $\sigma_{s}$.
This is due to the assumption of uncorrelated density distributions.
\subsection{Examples}
In Table \ref{tab:02} we give some examples of different colliders and
their luminosity and other relevant parameters.
\begin{table}[htb]
\begin{center}
\caption{Example of different colliders. We show the energy, luminosity, beam sizes and interaction rate for a~comparison.}\label{tab:02}
\begin{tabular}{|l|c|c|c|c|c|} \hline \hline
            & \textbf{Energy} & \textbf{$\cal{L}$} & \textbf{rate}  & \textbf{$\sigma{x}/\sigma_{y}$} & \textbf{Particles} \\
            & \textbf{(GeV)}    &  \textbf{cm$^{-2}$s$^{-1}$} & \textbf{s$^{-1}$} & \textbf{$\mu$m/$\mu$m} & \textbf{per bunch} \\ \hline
 SPS (p$\bar{p}$)  & 315x315    &   6 10$^{30}$   & 4 10$^{5}$ &  60/30 & $\approx$ 10 10$^{10}$ \\
 Tevatron (p$\bar{p}$)  & 1000x1000    &   50 10$^{30}$  & 4 10$^{6}$  &  30/30 & $\approx$ 30/8 10$^{10}$ \\
 HERA (e$^{+}$p)  & 30x920 &   40 10$^{30}$ & 40  &  250/50 & $\approx$ 3/7 10$^{10}$ \\
 LHC (pp)         & 7000x7000     &  10000 10$^{30}$& 10$^{9}$      &  17/17 & 11 10$^{10}$ \\
 LEP (e$^{+}$e$^{-}$) & 105x105  &   100 10$^{30}$ &  $\leq$ 1 &  200/2 & $\approx$ 5 10$^{11}$ \\
 PEP (e$^{+}$e$^{-}$) & 9x3  &  3000 10$^{30}$ &   NA  &  150/5 & $\approx$ 2/6 10$^{10}$ \\
 KEKB (e$^{+}$e$^{-}$) & 8x3.5  &  10000 10$^{30}$ &   NA  &  77/2 & $\approx$ 1.3/1.6 10$^{10}$ \\
\hline \hline
\end{tabular}
\end{center}
\end{table}
One may notice the very different interaction rate, in particular between 
hadron colliders and high energy lepton colliders.
This is due to the small total cross section of e$^{+}$e$^{-}$ interactions.
Furthermore, since so-called B-factories such as PEP and KEKB operate near or
on resonances, the interaction rate varies very strongly with the precise
energy. Therefore we write the term NA in the table.

\subsection{Additional complications in real machines}
So far we have assumed ideal head-on collisions
of bunches where the particle densities in the three
dimensions are uncorrelated.
In practice, we have to include additional effects in our
computations, some of the most important are:
\begin{itemize}
\item[-]Crossing angle                                               
\item[-]Collision offset (wanted or unwanted)
\item[-]Hour glass effect.                                                                    
\end{itemize}

Crossing angles are often used to avoid unwanted collisions
in machines with many bunches (e.g. LHC, CESR, KEKB).
Such crossing angles can have important consequences for
beam-beam effects \cite{beambeam} but also affect the
luminosity. 
When beams do not collide exactly head-on but with a small
offset, the~luminosity is reduced. 
Such an offset can be wanted (e.g. to reduce luminosity or during
measurements) or unwanted, for example as a result of
beam-beam effects \cite{beambeam}.
The so-called hourglass effect is a~geometrical
effect which includes a dependence of the transverse beam sizes
on the longitudinal position and therefore violates our
previous assumption of uncorrelated particles densities.
When the beam profiles deviate from a Gaussian function, we
may have to apply some correction factors and when the dispersion
at the interaction point is not zero, the effective beam sizes
are increased, leading to a smaller luminosity.
In case of optical imperfections the collision point may not
be at the minimum of the betatron function $\beta^{*}$, i.e.
at the waist, but slightly displaced with implications for the
effective beam sizes.
Some of the~most important of these additional effects 
we shall investigate in the following sections.

\subsubsection{Crossing angles}
A very prominent collider with a crossing angle is the
LHC presently under construction at CERN.
In the~LHC one has almost 3000 closely spaced bunches and to
avoid numerous unwanted interactions, the~two beams collide
at a total crossing angle of around $\approx$ 300 $\mu$rad.
\begin{figure}[ht]
  \centering
  \includegraphics[width=0.5\textwidth]{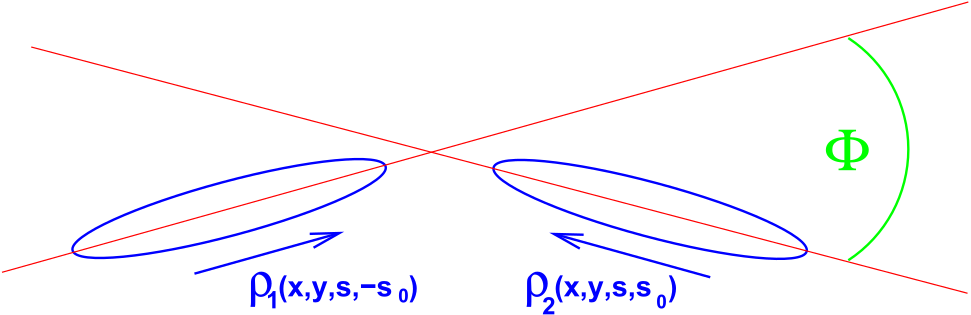}
  \caption{Schematic view of two bunches colliding at a finite crossing angle.}
  \label{fig:COL-WH03}
\end{figure} 

\Fref[b]{fig:COL-WH03} shows a schematic illustration of the collision region.
Colliders with unbunched, i.e. coasting beams such as the ISR need
a sizeable crossing angle to confine the interaction 
region (e.g. $\approx$ 18$^{o}$ at the ISR).
~

In the following we shall assume without loss of generality
that the crossing angle is in the horizontal plane.

\begin{figure}[ht]
  \centering
  \includegraphics[width=0.5\textwidth]{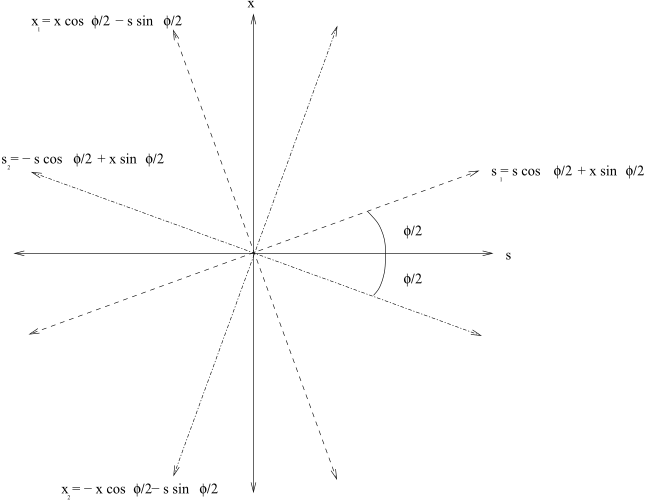}
  \caption{Rotated reference system for collisions at a finite crossing angle.}
  \label{fig:COL-WH04}
\end{figure} 

The overlap integrals are evaluated in the x and y coordinate system
and therefore we have to transform our bunches into the proper system.
The geometry of a collision at a crossing angle $\Phi$ 
is shown in \Fref{fig:COL-WH04}.
To make the treatment more symmetric, we have assumed that the total
crossing angle is made up by two rotations $\Phi/2$ and $-\Phi/2$ 
each of the two beams in the x-s plane (see \Fref{fig:COL-WH04}).

To compute the integral we have to transform $x$ and $s$ to 
new, rotated coordinates which are now different for the two beams:
\begin{eqnarray}\label{eq:018}
x_{1} = x\cos\frac{\phi}{2} - s\sin\frac{\phi}{2},~~~~~~~~
s_{1} = s\cos\frac{\phi}{2} + x\sin\frac{\phi}{2}  \\
x_{2} = x\cos\frac{\phi}{2} + s\sin\frac{\phi}{2},~~~~~~~~
s_{2} = s\cos\frac{\phi}{2} - x\sin\frac{\phi}{2}.
\end{eqnarray}
The overlap integral becomes:
\begin{eqnarray}\label{eq:019}
{\cal L} = {{2\cos^{2}\frac{\phi}{2}}}N_{1}N_{2}fN_{b}
\displaystyle{\int\int\int\int_{-\infty}^{+\infty}}
\rho_{1x}(x_{1})\rho_{1y}(y_{1})\rho_{1s}(s_{1}-s_{0})
\rho_{2x}(x_{2})\rho_{2y}(y_{2})\rho_{2s}(s_{2}+s_{0}) dxdydsds_{0}.
\end{eqnarray}
The factor $2\cos^{2}\frac{\phi}{2}$ is the kinematic factor
when the two velocities of the bunches are not collinear (from Eq.~(\ref{eq:009})).
~\\
After the integration over $y$ and $s_{0}$, using the formula:
\begin{eqnarray}\label{eq:020}
\displaystyle{\int^{+\infty}_{-\infty} e^{-(a t^{2} + b t + c)} dt~~~=~~~\sqrt{\pi/a} \cdot e^{\frac{b^{2} - a c}{a}}}
\end{eqnarray}
we get:
\begin{eqnarray}\label{eq:021}
{\cal L} = \frac{N_{1}N_{2}fN_{b}}{8\pi^{2}\sigma_{s}\sigma_{x}^{2}\sigma_{y}}{{2\cos^{2}\frac{\phi}{2}}}
\displaystyle{\int\int 
e^{-\frac{x^{2}\cos^{2}(\phi/2) + s^{2}\sin^{2}(\phi/2)}{\sigma_{x}^{2}}}}
e^{-\frac{x^{2}\sin^{2}(\phi/2) + s^{2}\cos^{2}(\phi/2)}{\sigma_{s}^{2}}} ~~~dxds.
\end{eqnarray}
We make the following approximations:
since both $x$ and sin($\phi/2$) are small, we drop all terms $\sigma^{k}_{x}sin^{l}(\phi/2$) or $x^{k}sin^{l}(\phi/2$) for all k+l$~\geq~4$
and approximate sin($\phi/2$) $\approx$ tan($\phi/2$) by $\phi/2$.
After the final integrations we get for the luminosity an expression
of the form: 
\begin{equation}\label{eq:022}
{\cal L} = \displaystyle{\frac{N_{1}N_{2}fN_{b}}
{4\pi\sigma_{x}\sigma_{y}}~\cdot~{S}}.
\end{equation}
This looks exactly like the well known formula we have derived already,
except for the additional factor
$S$, the so-called luminosity reduction factor which can be written as:
\begin{equation}\label{eq:023}
{S = \displaystyle{
\frac{1}{\sqrt{1+(\frac{\sigma_{x}}{\sigma_{s}}\tan\frac{\phi}{2})^{2}}}
\frac{1}{\sqrt{1+(\frac{\sigma_{s}}{\sigma_{x}}\tan\frac{\phi}{2})^{2}}}}}.
\end{equation}
For small crossing angles and $\sigma_{s} \gg \sigma_{x,y}$ we 
can simplify the formula to:
\begin{equation}\label{eq:024}
{S =
\displaystyle{\frac{1}{\sqrt{1+(\frac{\sigma_{s}}{\sigma_{x}} \tan\frac{\phi}{2})^{2}}}}~\approx~
\displaystyle{\frac{1}{\sqrt{1+(\frac{\sigma_{s}}{\sigma_{x}} \frac{\phi}{2})^{2}}}}}.
\end{equation}
~\\
A popular interpretation of this result is to consider it a correction
to the beam size and to introduce an~"effective beam size" like:
\begin{equation}\label{eq:025}
\displaystyle{\sigma_{eff}~=~\sigma~\cdot~{\sqrt{1+(\frac{\sigma_{s}}{\sigma_{x}} \frac{\phi}{2})^{2}}}}.
\end{equation}
The effective beam size can then be used in the standard formula
for the beam size in the crossing plane.
This concept of an effective beam size is interesting because it
also applies to the calculation of beam-beam effects of bunched
beams with a crossing angle \cite{jja1}.

As an example we use the parameters of the LHC.
The number of particles per bunch is 1.15 10$^{11}$, the beam sizes
in the two planes $\approx$ 16.7~$\mu$m, the bunch 
length $\sigma_{s}$ = 7.7~cm and the
total crossing angle $\Phi~=~$285~$\mu$rad.
With the revolution frequency of 11.245 kHz and 2808 bunches, we
get for the head-on luminosity 1.2~10$^{34}$~cm$^{-2}$s$^{-1}$.
For the luminosity reduction factor $S$ we get 0.835 and the final
LHC luminosity with a crossing angle becomes $\approx$~1.0~10$^{34}$~cm$^{-2}$s$^{-1}$.

\subsubsection{Crossing angles and offset beams}
A modification of the previous scheme is needed when the beams do not
collide head-on, but with a~small transverse offset.
In order to be general, we shall treat the case with crossing angle
and offsets.
\Fref[b]{fig:COL-WH045} shows the modified geometry when we have the
same crossing angle as before, but beam 1 is displaced by $d_{1}$ and
beam 2 is displaced by $d_{2}$ with respect to their reference orbits.
\begin{figure}[ht]
\label{bnote2}
  \centering
  \includegraphics[width=0.5\textwidth]{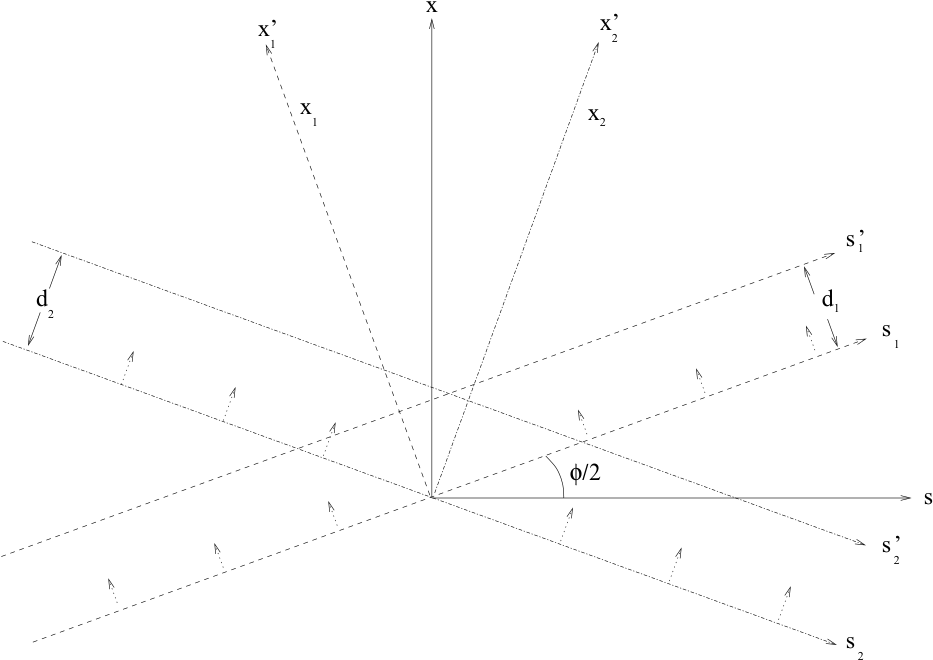}
  \caption{Schematic view of two bunches colliding at a finite crossing angle and an offset between the two beams.}
  \label{fig:COL-WH045}
\end{figure}

~~\\
The coordinate transformations are now:
\begin{eqnarray}\label{eq:026}
x_{1} = {{d_{1}}} + x\cos\frac{\phi}{2} - s\sin\frac{\phi}{2},~~~~~~~~
s_{1} = s\cos\frac{\phi}{2} + x\sin\frac{\phi}{2} \\
x_{2} = {{d_{2}}} + x\cos\frac{\phi}{2} + s\sin\frac{\phi}{2},~~~~~~~~
s_{2} = s\cos\frac{\phi}{2} - x\sin\frac{\phi}{2}.
\end{eqnarray}
Following the previous strategy and approximations for the integration,
we get after integrating $y$ and $s_{0}$:
\begin{eqnarray*}  
{\cal L} = \frac{N_{1}N_{2}fN_{b}}{8\pi^{2}\sigma_{s}\sigma_{x}^{2}\sigma_{y}}{{2\cos^{2}\frac{\phi}{2}}}
\displaystyle{\int\int 
e^{-\frac{x^{2}\cos^{2}(\phi/2) + s^{2}\sin^{2}(\phi/2)}{\sigma_{x}^{2}}}}
e^{-\frac{x^{2}\sin^{2}(\phi/2) + s^{2}\cos^{2}(\phi/2)}{\sigma_{s}^{2}}} \\
\end{eqnarray*}  
\begin{eqnarray}\label{eq:027}  
\displaystyle{
\times e^{-\frac{d_{1}^{2} + d_{2}^{2} + 2(d_{1} + d_{2})x\cos(\phi/2)
- 2(d_{2} - d_{1})s\sin(\phi/2)}
{2\sigma_{x}^{2}}}dxds}.
\end{eqnarray}  
~\\
After the integration over x we obtain: 
\begin{equation}\label{eq:028}
{\cal L} = \frac{N_{1}N_{2}fN_{b}}{8\pi^{\frac{3}{2}}\sigma_{s}}
{{2\cos\frac{\phi}{2}}}\int_{-\infty}^{+\infty}W
\frac{e^{-(As^2 + 2Bs)}}{\sigma_{x}\sigma_{y}}ds
\end{equation}
with: 
\begin{equation}\label{eq:029}
A = \frac{\sin^{2}\frac{\phi}{2}}{\sigma_{x}^{2}} +
\frac{\cos^{2}\frac{\phi}{2}}{\sigma_{s}^{2}} ,\ \ \ 
B = \frac{(d_{2} - d_{1})\sin(\phi/2)}{2\sigma_{x}^{2}} 
\end{equation}
and 
\begin{equation}\label{eq:030}
W = e^{-\frac{1}{4\sigma_{x}^{2}}(d_{2} - d_{1})^{2}}.
\end{equation}
We can re-write the luminosity with three correction factors:
\begin{equation}\label{eq:031}
{\cal L} = \frac{N_{1}N_{2}fN_{b}}{4\pi\sigma_{x}\sigma_{y}}\cdot{{W}}
\cdot{{e^{\frac{B^{2}}{A}}}}\cdot{{S}}. 
\end{equation}
This factorization enlightens the different contributions
and allows straightforward calculations.
The last factor $S$ is the already calculated luminosity reduction
factor for a crossing angle.
One factor $W$ reduces the luminosity in the presence of 
beam offsets and the factor $e^{\frac{B^{2}}{A}}$ is only
present when we have a crossing angle and offsets simultaneously.
\subsubsection{Hourglass effect}
So far we have assumed uncorrelated beam density functions
in the transverse and longitudinal planes.
In particular, we have assumed that the transverse beam sizes
are constant over the whole collision regions.
However, since the $\beta$-functions have their minima at the
collision point and increase with the distance this is not
always a good approximation.
\begin{figure}[ht]
  \centering
  \includegraphics[width=0.5\textwidth]{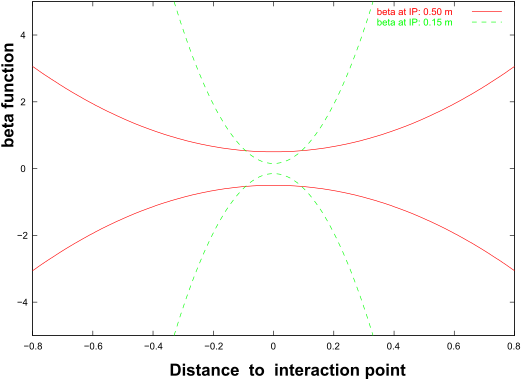}
  \caption{Schematic illustration of the hourglass effect. $\beta$(s) is plotted for two different values of $\beta^{*}$.}
  \label{fig:COL-WH05}
\end{figure}

In a low-$\beta$ region the $\beta$-function varies with the
distance $s$ to the minimum like:
\begin{equation}\label{eq:032}
\beta(s) = \beta^{*}({1+\left(\frac{s}{\beta^{*}}\right)^{2}})
\end{equation}
and therefore the beam size $\sigma~=~\sqrt{\beta(s) \cdot \epsilon}$
increases approximately linearly with the distance to the~interaction point.
This is schematically shown in \Fref{fig:COL-WH05} where the functions $\beta$(s)
are shown for two different values of $\beta^{*}$ (0.50~m and 0.15~m).
Because of the shape of the $\beta$(s) function this effect is called
the hourglass effect.
It is especially important when the $\beta$(s) function at the interaction
point approaches the bunch length $\sigma_{s}$ (\Fref{fig:COL-WH05}) and
not all particles collide at the minimum of the transverse beam size,
therefore reducing the luminosity.
Other effects such as a coupling between the transverse and longitudinal
planes are ignored in this discussion.

In our formulae we have to replace $\sigma$ by $\sigma$(s)
and get a more general expression for the luminosity:
\begin{equation}\label{eq:033}
\ \ \displaystyle{{\cal L} = \left(\frac{N_{1}N_{2}fN_{b}}
{8\pi\sigma_{x}^{*}\sigma_{y}^{*}}\right)
{\frac{{{2\cos\frac{\phi}{2}}}}{\sqrt{\pi}\sigma_{s}}\int_{-\infty}^{+\infty}
\frac{e^{-s^{2}A}}{{{1+(\frac{s}{\beta^{*}})^{2}}}}ds}}
\end{equation}
with
\begin{equation}\label{eq:034}
\ \ {\displaystyle{
A = \frac{\sin^{2}\frac{\phi}{2}}{(\sigma_{x}^{*})^{2}{{[1+(\frac{s}{\beta^{*}})^{2}]}}}} +
\frac{\cos^{2}\frac{\phi}{2}}{\sigma_{s}^{2}}}.  \ \ \ 
\end{equation}

Usually it is difficult to compute this integral analytically
and it has to be evaluated by numerical integration.

To estimate the importance and relevance of this effect,
we shall use the parameters of the LHC,
i.e. $N_{1} = N_{2} = 1.15\times 10^{11}$ particles/bunch,
2808 bunches per beam,
a revolution frequency of
f~=~11.2455~kHz, and a crossing angle of $\phi = 285~\mu$rad.           
The nominal $\beta$-functions at the interaction point are
$\beta_{x}^{*} = \beta_{y}^{*} = 0.55~$m, leading to beam sizes of
$\sigma_{x}^{*} = \sigma_{y}^{*} = 16.7 \ \mu$m,  and we use 
a r.m.s. bunch length of $\sigma_{s}$ = 7.7~cm.

In the simplest case of a head on collision we get for the luminosity
${\cal L} = 1.200\times10^{34} \ \mathrm{cm}^{-2}\mathrm{s}^{-1}$.
~\\
The effect of the crossing angle we can estimate by the evaluation
of the factor $S$ and get:
${\cal L} = 1.000\times10^{34} \ \mathrm{cm}^{-2}\mathrm{s}^{-1}$.

When we further include the hourglass effect we get:
${\cal L} = 0.993\times10^{34} \ \mathrm{cm}^{-2}\mathrm{s}^{-1}$.

While the effect of the crossing angle is very sizeable (S = 0.835),
the further reduction by the~hourglass effect is small, at least
for the nominal LHC parameters. 
For smaller $\beta$-functions at the interaction point this may 
not be the case.
\newpage
\section{Other luminosity issues}
There are further issues related to the luminosity which
are important for the experiments, such as:
\begin{itemize}
\item Integrated luminosity
\item Time structure of interactions
\item Space structure of interactions.
\end{itemize}
The geometry of the interaction regions as well as some basic
parameters entering the standard luminosity formulae
are very important for the above issues and may need
reconsideration to trade off between the~different requirements.
\subsection{Integrated luminosity}
\subsubsection{Definition of integrated luminosity}
The maximum luminosity, and therefore the instantaneous number
of interactions per second, is very important, but the
final figure of merit is the so-called integrated luminosity:
\begin{equation}\label{eq:038}
{\cal L}_{\mathrm{int}} = \displaystyle{
\int_{0}^{T}{\cal L}(t')dt'}   
\end{equation}
because it directly relates to the number of observed events:
\begin{equation}\label{eq:039}
{\cal L}_{\mathrm{int}} \cdot \sigma_{p} = {\mathrm{number~of~events~of~interest}}.
\end{equation}
The integral is taken over the sensitive time, i.e. excluding
possible dead time.
For an evaluation one needs a realistic model for the decay 
of the luminosity with time.
Different possibilities exist and
usually one assumes some behaviour (e.g. exponential)
with a given lifetime ${\tau}$:
\begin{equation}\label{eq:040}
 {\cal{L}}(t) \longrightarrow {\cal{L}}_{0}\exp{\left(-\frac{t}{\tau}\right)}. 
\end{equation}
Contributions to this life time we have from the decay of beam intensity
with time, the growth of the~transverse emittance, increase of
the bunch length etc.
The advantage of assuming an exponential decay is that the contributions
from different processes can be easily added.
The differences between the~different models is very small in practice.
\subsubsection{Optimization of integrated luminosity}
\begin{figure}[ht]
  \centering
  \includegraphics[width=0.7\textwidth]{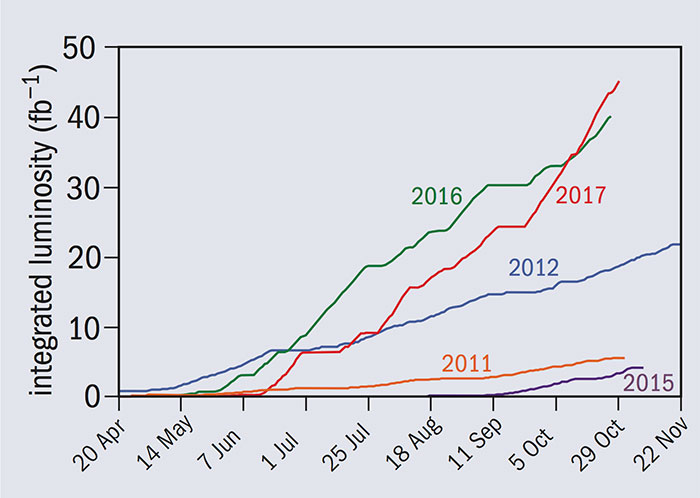}
  \caption{Typical performance curve of a collider: The LHC integrated luminosity for several years of running [taken from CERN courier].}
  \label{fig:intlum}
\end{figure} 
The aim of the operation of a collider must be to optimize
the integrated luminosity.
Two parts of the~operation must be distinguished: the luminosity
run with a lifetime $\tau$ and the preparation time between
two luminosity runs t$_{p}$.
The optimization problem is very similar to the 
challenge of a formula 1 racing team:
the length of running with decreasing performance (slowing down with
ageing tires)
and the~time needed to restore the performance (changing tires).
The best strategy should minimize the overall time needed.

In principle, the knowledge of the preparation time allows 
an optimization of ${\cal L}_{\mathrm{int}}$.
~\\             
If we assume an exponential decay of the luminosity ${\cal{L}}(t) = {\cal{L}}_{0} \cdot e^{t/\tau}$
we want to maximize the average luminosity $<{\cal{L}}>$:      
\begin{equation}\label{eq:041}
<{\cal{L}}> = \frac{\int_{0}^{t_{r}} {\cal{L}}(t) dt}{t_{r} + t_{p}} = {\cal{L}}_{0} \cdot \tau \cdot \frac{1 - e^{-t_{r}/\tau}}{t_{r} + t_{p}}. 
\end{equation} 
Here $t_{r}$ is the length of a luminosity run and $t_{p}$ the preparation
time between two runs.
Since $t_{r}$ is a~"free" parameter, i.e. can be chosen
by the operation crew, we can optimize this expression and 
get a~(theoretical) maximum for: 
\begin{equation}\label{eq:042}
t_{r} \approx \tau \cdot {\mathrm{ln}}(1 + \sqrt{2t_{p}/\tau} + t_{p}/\tau).
\end{equation}
Assuming some parameters for the LHC \cite{ob1}: \\
$t_{p}\approx$~10h, $\tau \approx$~15h, we get: $\Rightarrow t_{r} \approx$~15h .
\Fref[b]{fig:intlum} shows as an example a typical performance curve as it has been published for the CERN LHC collider. For every year of running the integrated luminosity is shown. Taking the value at the end of the year and multiplying it with the cross-section of the interaction of interest, one gets easily the number of produced events.

\subsection{Luminous region and space structure of luminosity}
In addition to the number of events, the space structure is important
for the design and running of a~particle physics experiment.
The questions we asked are therefore:
\begin{itemize}
\item[-] What is the density distribution of interaction vertices ?
\item[-] Which fraction of collisions occur {{$\pm$ s}} from the interaction point ?
\end{itemize}
The answers depend on beam properties such as
$\sigma_{x}$, $\sigma_{y}$, and $\sigma_{s}$
but also on the crossing angle $\phi$.
\begin{figure}[ht]
  \centering
  \includegraphics[width=0.6\textwidth]{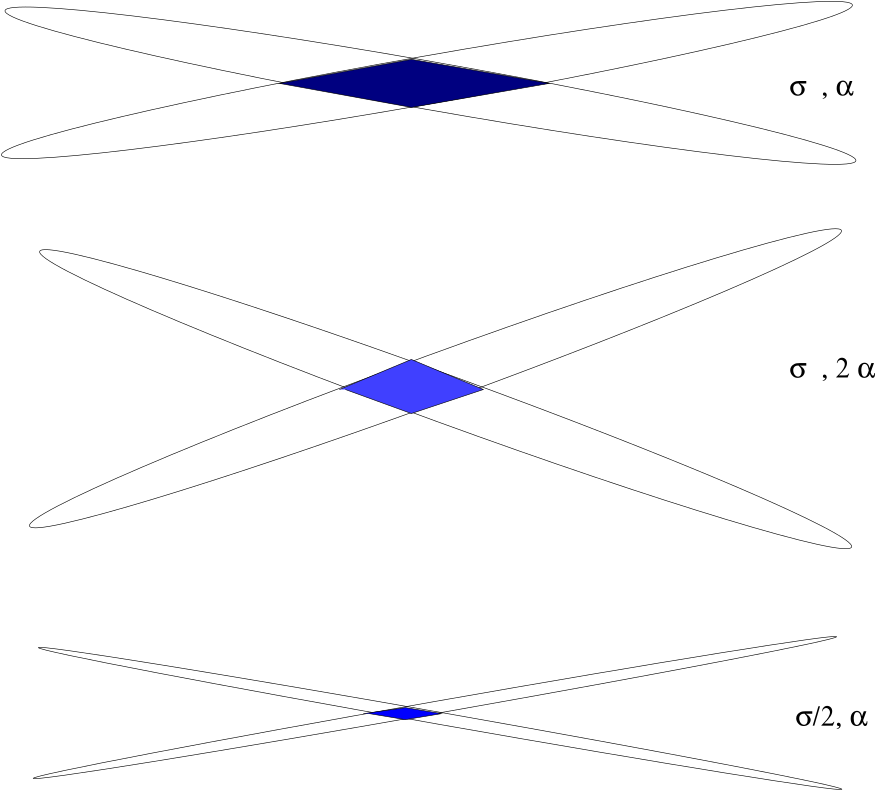}
  \caption{Schematic illustration of luminous regions.}
  \label{fig:COL-WH06}
\end{figure}                         

This is very schematically indicated by the overlap regions in
\Fref{fig:COL-WH06}.
Depending of the beam and machine parameters, this region can be
very different, with important consequences for e.g. trigger system
or pattern recognition.
We evaluate:
\begin{equation}\label{eq:043}
{\cal L}_{0} = \int_{-\infty}^{+\infty}{\cal L}(s')ds' \longrightarrow {\cal L}({{s}}) = \int_{{{-s}}}^{{{+s}}}{\cal L}(s')ds' 
\end{equation}
and get:
\begin{equation}\label{eq:044}
{\cal L}(s) =
\left(\frac{N_{1}N_{2}fN_{b}}{8\pi\sigma_{x}^{*}\sigma_{y}^{*}}\right)
\frac{2\cos\frac{\phi}{2}}{\sqrt{\pi}\sigma_{s}}
\ \sqrt{\frac{\pi}{A}} \ \mathrm{erf}\left(\sqrt{A} \ s \right).
\end{equation}
For the integrated luminosity this becomes:
\begin{equation}\label{eq:045}
{\cal L}_{\mathrm{int}}(s) = \displaystyle{
\int_{0}^{T}\int_{-s}^{+s}{\cal L}(s',t)ds'dt}.   
\end{equation}

In order to evaluate this numerically, we use again LHC nominal
parameters as above.
The results of our calculations are shown in Tab.\ref{tab:03}.
\begin{table}[htb]
\centering
\caption{Percentage of visible luminosity as a function of distance to interaction point.}\label{tab:03}
\begin{tabular}{|c|c|}  \hline \hline
            &     \\
  \textbf{Integration range}  & \textbf{Percentage of luminosity} \\
            &     \\
\hline
            &         \\
 s~=~$\pm$~12~cm      &   1.000          \\
            &     \\
 s~=~$\pm$~ 8~cm      &   0.950          \\
            &     \\
 s~=~$\pm$~ 7~cm      &   0.900          \\
            &     \\
 s~=~$\pm$~ 6~cm      &   0.850          \\
            &     \\
 s~=~$\pm$~ 5.5~cm      &   0.800          \\
            &     \\
\hline \hline
\end{tabular}
\end{table}
While practically all luminosity is seen at a distance of $\pm$ 12~cm from
the interaction point, about 20\% is lost when only a region of
$\pm$ 5.5~cm is covered by the detector or the software.
This does strongly depend on the crossing angle.
A detailed examination of this property was done in \cite{bruno1}.
\subsection{Time structure of luminosity}
In addition to the space structure, the time structure of the
interactions is an important input for the setup of an experiment
and even on the possible physics that can be studied.

In the LHC the bunches cross every 25~ns and it can be calculated easily
that for proton-proton collisions one has to expect $\approx$~20 
simultaneous interactions per bunch crossing.
They must be digested by the detectors before the
next bunch crossing occurs, a non trivial task for the
experiments.
Some physics studies that cannot be done with such an event pile up
may require to run at a lower luminosity.

\section{Collider examples of "yesterday, today and tomorrow"}

In the last section we highlight some particle colliders, in order to draw the line from first steps in the~1960's to future CERN projects.

\subsection{AdA - first $e^+ e^-$ collider ever (1961-1964)}
\begin{figure}[ht]
  \centering
  \includegraphics[width=0.5\textwidth]{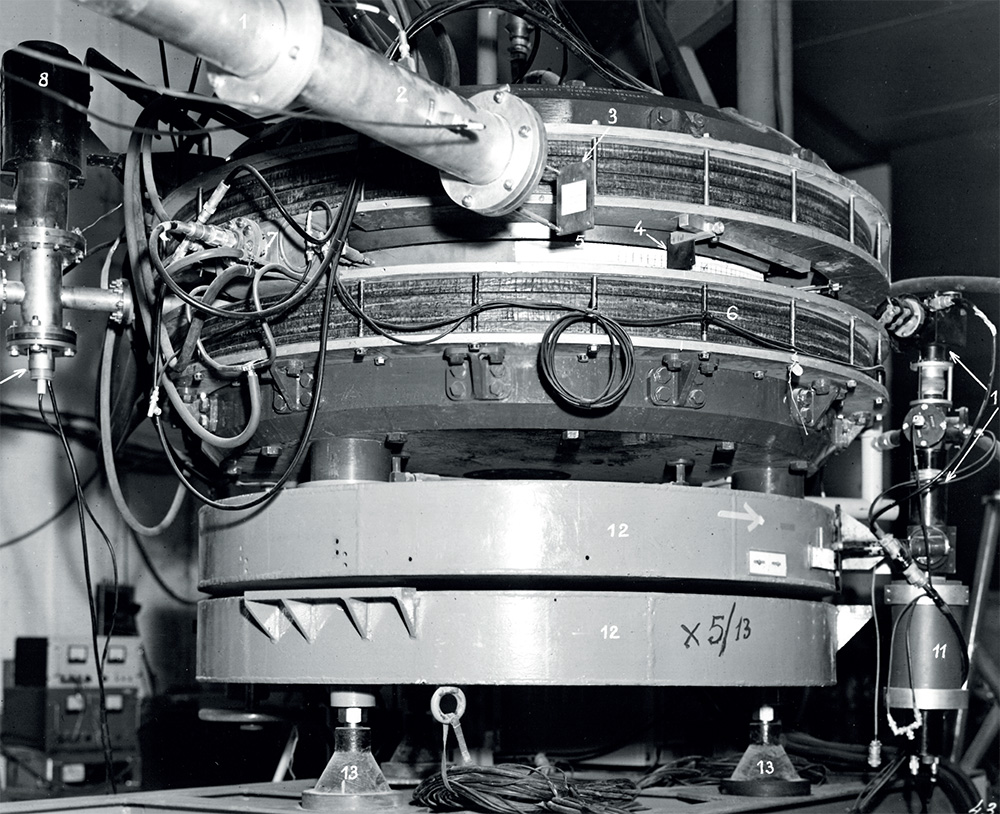}
  \caption{Photo of the first $e^+ e^-$ collider AdA in Frascati (taken from CERN courier). }
  \label{fig:COL-AdA}
\end{figure}
In the year 1961 the Austrian physicist Bruno Touschek proposed the idea of colliding beams and proposed a demonstrator installation at the LNF (Frascati National Laboratory) in Frascati. At that time many people working on fixed target experiments disliked the idea of replacing the solid target with another counter-rotating beam. 
AdA went even one step further replacing the second beam with a beam of antimatter in order to use the same magnetic system.

After the machines construction, it was operated from 1961 to 1964, by the National Institute of Nuclear Physics, in Frascati, Italy. However, in 1962, the machine was relocated to the Laboratoire de l’Accelerateur Lineaire in Orsay, France, where it was used for an additional four years alongside the~laboratory's powerful particle injector.
 
Towards the end of 1963, AdA's first electron-positron collisions were recorded. Then, the machine was operated a few more years for further successful and decisive tests before dismantling. AdA was never used to collect physics data. Instead, it was a testing ground for a breed of machines that was to change the course of particle physics in the following decades.

\subsection{Further milestones in collider developments}

For many years people were sceptical about the possibilities to collide two counter-rotating beams. The~worries were that the forces at the interaction moment were so large that the two beams would disintegrate after the collisions (in particular in the case of bunched beams).
All first collider attempts were with electrons and positrons, the first proton-proton collider was the ISR (:= intersection storage rings) at CERN, which used unbunched beams and as consequence large crossing angles. Light ions (alpha, deuterium...) were also collided in the ISR.
Also at the ISR first collisions of protons and antiprotons were made in the year 1981 including first cooling experiments in order to reduce the emittance of the~antiproton beam (and hence increase the luminosity of the collisions).

The developments at the ISR have led to the construction of the antiproton accumulator (AA) at CERN paving the way to use the existing SPS fixed target synchrotron as proton-antiproton collider. This happened in the early 1980's at the same time when CERN was preparing for the construction and operation of LEP.

Significant work on increasing the performance of lepton colliders has been done at CESR in Ithaca (NY, United States of America) (= Cornell electron storage ring). Detailed studies on beam-beam forces were carried out there plus the development of a so called "Pretzel" Scheme, in which the~electrons and positrons were sent on largely different orbits in the accelerator arcs avoiding parasitic beam crossings in the arc. With a Pretzel scheme a collider can accelerate and store many bunches of particle-antiparticle beams using a single beam-pipe.

\subsection{LEP - $e^+ e^-$ collider (1989 - 2000)}

The direct discovery of the W and Z bosons at the CERN SPS proton-antiproton collider in the year 1982 has defined the energy scale for the next generation of $e^+ e^-$ colliders, for which the principle task was to study with high precision the Z-boson (LEP and SLC) and the charge W bosons (LEP-phase II).
In Europe CERN set out to construct a circular collider with an energy reach of 200 GeV, whereas in the~United states at Stanford one opted for the more challenging solution of constructing a linear collider (SLC = Stanford linear collider, next section).

LEP was operated in two phases, first at a cms energy of about 91 GeV in order to mass produce Z bosons, later above 200 GeV in order to produce the charged W bosons in pairs.

In order to reach 100 GeV beam energy a 27 km circumference tunnel was constructed.
The~collider was filled with only 4 bunches of both beams producing collisions in four interaction points. 

The key technology in order to compensate for the synchrotron radiation loss was an Rf system composed of conventional normal conducting Rf cavities plus a large number of Nb sputtered superconducting cavities.
\begin{figure}[ht]
  \centering
  \includegraphics[width=0.9\textwidth]{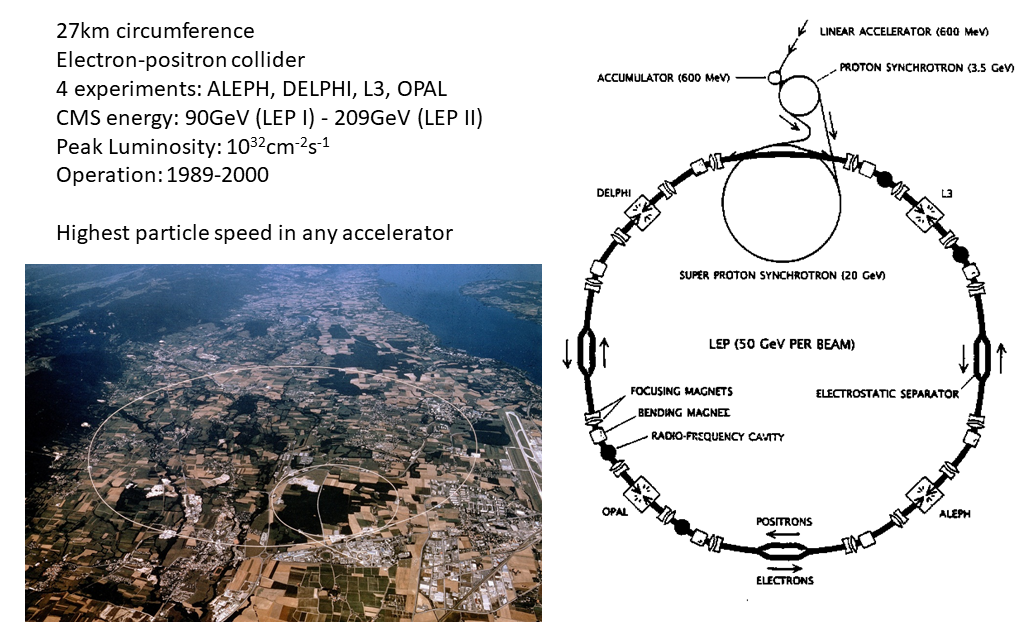}
  \caption{Summary sheet of the CERN LEP $e^+ e^-$ collider (1989 - 2000) }
  \label{fig:COL-LEP}
\end{figure}

\subsection{SLC - $e^+ e^-$ linear collider (1988 - 1998)}

Almost parallel to the efforts at CERN people at SLAC launched the effort to design, build and operate the first linear collider SLC. The accelerator was unique in design since it used the same 3.2 km long linac in order to accelerate the electrons and the positrons in following buckets of the RF system (with opposite sign). At the end of the linac a dipole magnet has split both beams into two $180^0$ return arcs in order to have head-on collisions.
The synchrotron radiation losses on one half turn created an energy spectrum of the colliding beams.
Enabling technology was apart from the linac RF system (normal-conducting) the development of damping rings, in which the emittance of the beams was reduced with the help of synchrotron radiation before entering the main linac.
Also, and contrary to LEP, the beams could be polarized in collision, making some unique physics experiment possible.

\begin{figure}[ht]
  \centering
  \includegraphics[width=0.75\textwidth]{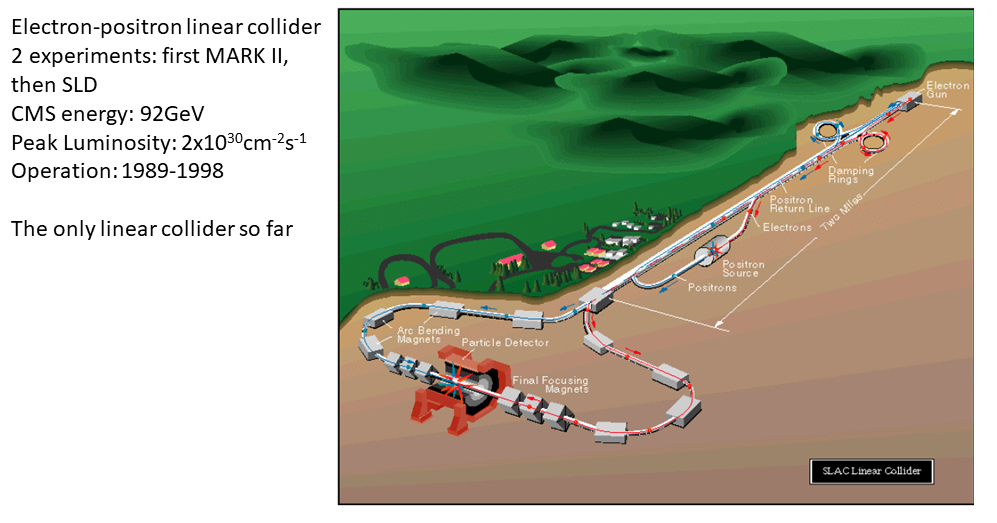}
  \caption{Summary sheet of the SLAC SLC $e^+ e^-$ collider (1988 - 1998) }
  \label{fig:COL-SLC}
\end{figure}

\subsection{LHC - pp collisions (2008 - today)}

Already in the 1990's CERN decided to reuse after the successful running of LEP the tunnel infrastructure in order to build a superconducting two ring proton-proton collider: the Large Hadron Collider LHC. It started operation in the year 2008 and it is as of today still running.
The physics objective in general are searches for high mass particles, such as the Higgs boson (discovered at the LHC in the year 2012) and possibly particles beyond the Standard Model (BSM-physics).

In the case of the LHC the enabling technology are superconducting dipoles and quadrupoles. The~dipoles have a field strength above 8 Tesla.

The four main physics experiments are ATLAS, CMS, ALICE and LHCb. The LHC delivers peak luminosities above $10^{34}cm^{-2}s^{-1}$ at a cms energy of 14 TeV.

\subsection{CLIC - $e^+ e^-$ future CERN linear collider}

Already for more than 30 years CERN pursues also design studies for a linear $e^+ e^-$ collider called CLIC. This collider could be built in several phases with different physics objectives. In the last phase an energy reach of 3 TeV cms energy is envisaged, opening the possibility for detailed studies of particles beyond the Standard Model.

The most important technology item is the up to 50 km long normal conducting linac, which aims at an accelerating record gradient of 100 MV/m. In order to deliver the enormous peak RF power for such a linac, a so called drive beam scheme is envisaged, by which a second low energy , but high intensity beam travelling parallel to the main beam transfers its energy for acceleration.
The  conceptional design report can be found under \cite{bib:COL-CLIC}.
\begin{figure}[ht]
  \centering
  \includegraphics[width=0.8\textwidth]{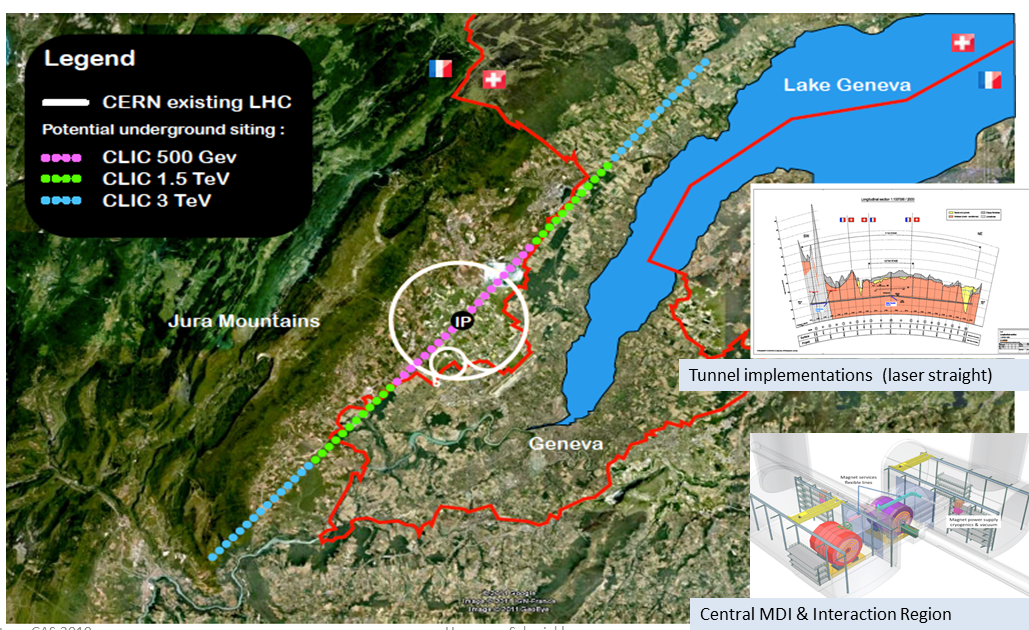}
  \caption{Possible implementation of the CLIC in the Geneva area}
  \label{fig:COL-CLIC}
\end{figure} 
\subsection{FCC - future CERN circular collider}

In the coming decades CERN sets out to rewrite the LEP/LHC history with another large circular collider complex (FCC = Future CERN Collider). A tunnel circumference of 100 km is envisaged.
\begin{figure}[ht]
  \centering
  \includegraphics[width=0.5\textwidth]{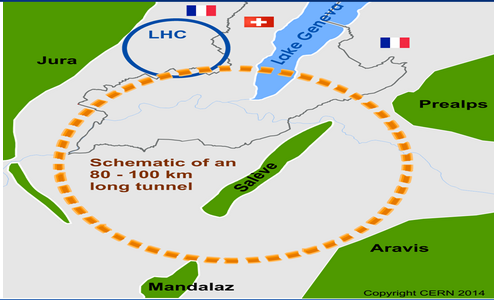}
  \caption{Possible implementation of the FCC  in the Geneva area}
  \label{fig:COL-FCC}
\end{figure} 

Again the project would start with an $e^+ e^-$ collider, which would also run at the previous LEP energies, but with more than 100 times the luminosity.
The energy scale could be extended up to 360 GeV, which would allow to study the t-quark produced as $t\bar t$ system.
The enabling technology is predominantly large scale superconducting RF systems.

After operation of the lepton machine, the installation of a proton-proton collider is envisaged in the same tunnel. Here one would gain compared to the LHC almost a factor 4 in ring bending radius and another factor 2 in the dipole strength (aim: 16 Tesla). This would give a total cms energy of 100 TeV, which will put this collider forward as search engine for new particles. The conceptual design report can be found under \cite{bib:COL-FCC}.

\begin{figure}[ht]
  \centering
  \includegraphics[width=0.6\textwidth]{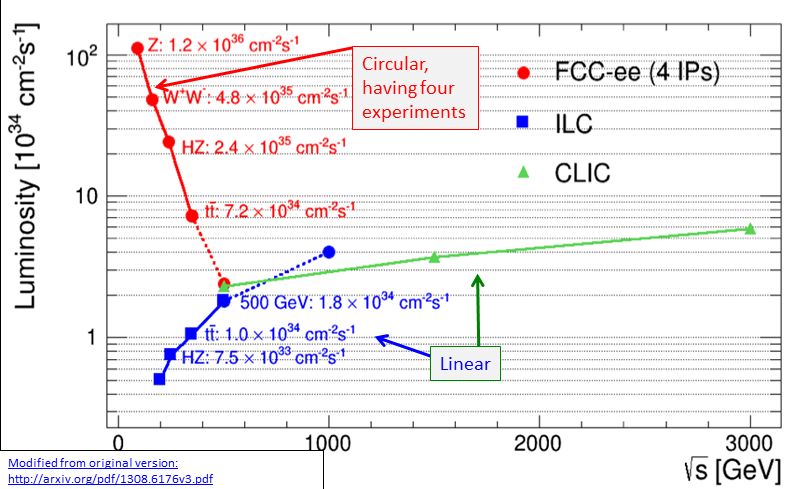}
  \caption{Luminosity-energy chart for the CERN options for future $e^+ e^-$ colliders}
  \label{fig:COL-FCClum}
\end{figure} 
\Fref[b]{fig:COL-FCClum} compares the linear and circular collider options at CERN in respect to energy reach and design luminosities.

\section{Acknowledgements}

I would like to thank Werner Herr (CERN, now EPFL Lausanne, Switzerland), who has given in the past a lecture on collider luminosity for letting me use a significant part of his material including the write-up of the luminosity calculations.
Some of the material has been elaborated by Bruno Moratori (now STFC Daresbury) during his fellowship at CERN.


\begin{thebibliography}{99}
\bibitem{bib:COL1} 
Frank W.~K.~Firk, Introduction to Relativistic Collisions.\\
\url{https://arxiv.org/ftp/arxiv/papers/1011/1011.1943.pdf}

\bibitem{bib:COL2}
C.~Biscari, (8/2002)., Accelerators R\& D., Proceedings of Science (journal).

\bibitem{bib:COL3}
Measurement of the mass of the Z boson and the energy calibration of LEP,\\
Working Group on LEP Energy, L.~Arnaudon et al., Phys.Lett.B 307 (1993) 187-193,\\
DOI: 10.1016/0370-2693(93)90210-9.

\bibitem{bib:COL4}
Muon Colliders, J.P.Delahaye et al.,\\
\url{https://arxiv.org/abs/1901.06150}.

\bibitem{bib:muon}
Muon Collider. A Path to the Future?,
D.Schulte et al., Proceedings of Science,
European Physical Society Conference on High Energy Physics - EPS-HEP2019, 10-17 July, 2019, Ghent, Belgium.
\url{https://indico.cern.ch/event/867138/attachments/1954116/3245304/Muon_Collider_EPS_2019.pdf}

\bibitem{bib:COL5}
Proton-Proton and Proton-Antiproton Colliders, W.Scandale (2014),
Rev.Accel.Sci.Tech.7 (2014) 9-33\\
DOI: 10.1142/S1793626814300023.

\bibitem{bib:COL6}
The proton laid bare, M.~Rayner (2019), CERN Courier.\\
\url{https://cerncourier.com/a/the-proton-laid-bare/}

\bibitem{bib:COL7}
RHIC project overview, M.Harrison et al., \\
Nucl.Instrum.Meth.A 499 (2003) 235-244, DOI: 10.1016/S0168-9002(02)01937-X.

\bibitem{bib:COL8}
\url{https://www.desy.de/research/facilities__projects/hera/index_eng.html}

\bibitem{bib:COL9}
Photon-Photon Collisions - PAst and Future, S.~J.~Brodsky (SLAC), 2003.\\
\url{https://www.slac.stanford.edu/pubs/slacpubs/11500/slac-pub-11581.pdf}

\bibitem{bib:COL10}
Advanced acceleration concepts, M.Ferrario, these proceedings.

\bibitem{bib:COL11}
Proceedings of the CAS course on "High Gradient Wakefield Accelerators", SESIMBRA (PO), 2019. \url{https://cas.web.cern.ch/schools/sesimbra-2019}

\bibitem{bib:COL-CLIC}
A Multi-TeV Linear Collider based on CLIC-Technology, CERN yellow report CERN-2012-007.\\
\url{https://project-clic-cdr.web.cern.ch/CDR_Volume1.pdf}

\bibitem{bib:COL-FCC}
\url{https://fcc-cdr.web.cern.ch/}

\bibitem{kinfact} C.~Moller, K. Danske Vidensk. Selsk. Mat.-Fys. Medd., {\bf{23}}, 1 (1945).
\bibitem{beambeam} W.~Herr, {\em{Beam-beam interactions}}, this school.

\bibitem{here} H.G.~Hereward, {\em{How good is the r.m.s. as a measure of the beam size ?}}, CERN/MPS/DL 69-15  (CERN, November 1969).

\bibitem{ob1} O. Br\"{u}ning, private communication.

\bibitem{jja1} J.E. Augustin, {\em{Space charge effects in e$^{+}$e$^{-}$ storage rings with beams crossing at an angle}}, Orsay, Note interne 35-69 (1969).

\bibitem{bruno1} B.~Muratori, {\em{Luminosity and luminous region calculations for the LHC}}, LHC Project Note 301 (2002).

\bibitem{optt} Review of Particle Physics, Vol. {\bf{15}}, Number 1-4, 213 (2000).

\end{thebibliography}
\end{document}